\documentclass[12pt]{article}
\usepackage{colortbl}

 \catcode`@=11
  
\begin{document}
\thispagestyle{empty}
\def\on#1#2{{\buildrel{\mkern2.5mu#1\mkern-2.5mu}\over{#2}}}
\catcode`@=11
\def\secteqno{\@addtoreset{equation}{section}%
\def\theequation{\thesection.\arabic{equation}}}
\catcode`@=12
\secteqno
\newcommand{\bea}{\begin{eqnarray}}
\newcommand{\eea}{\end{eqnarray}}
\newcommand{\bref}[1]{(\ref{#1})}
\newcommand{\nn}{\nonumber}\vfill
\hfill February 5, 2003\par
\hfill KEK-TH-868\null\par
\vskip 20mm
\begin{center}
{\Large\bf AdS and pp-wave D-particle superalgebras }\par
\vskip 6mm
\medskip
\vskip 10mm
{\large Machiko\ Hatsuda }\par
\medskip
{\it 
Theory Division,\ High Energy Accelerator Research Organization (KEK),\\
 Tsukuba,\ Ibaraki,\ 305-0801, Japan }\\
\medskip\medskip
{\small\sf E-mails:\ mhatsuda@post.kek.jp} 
\medskip
\end{center}
\vskip 10mm
\begin{abstract}
We derive
anticommutators of supercharges 
with a brane charge 
for
a D-particle
 in  AdS$_2\times$S$^2$ and  pp-wave backgrounds.
A coset $GL(2{\mid}2)/(GL(1))^4$ and its Penrose limit are
used with the supermatrix-valued coordinates
for the AdS and the pp-wave spaces respectively.
The brane charges have position dependence, 
and can be absorbed into bosonic generators by shift of momenta
which results in closure of the superalgebras.
\end{abstract} 
\par\par
\newpage
\setcounter{page}{1}
\parskip=7pt

\section{ Introduction}\par

Recently brane dynamics in curved backgrounds are
examined extensively
motivated by the AdS/CFT correspondence \cite{Malda} 
and its pp-wave limit \cite{jose} and \cite{BMN}.
The study of superstrings in AdS spaces has been started 
in earlier works \cite{MT,Ram,Zhou},
and further developed to pp-wave spaces \cite{Me1}
and to D-branes \cite{ske,ppex}. 
Branes are characterized by brane charges
 in superalgebras 
which are topological \cite{WitOl,azcTow}.
Superalgebras manifest
non-perturbative dualities  \cite{TowMS}
and present solitonic solutions
systematically \cite{Towdem}.
In general a superalgebra is uniquely obtained coordinate independently,
 while Killing spinor equations are coordinate dependent 
where there are several different coordinate systems in curved spaces.
So it will  give better outlook to examine 
superalgebras with brane charges in curved spaces also.
For a flat background $p$-brane charges are center \cite{TowMS},
and classified by non-trivial elements of the Chevalley-Eilenberg (CE) cohomology
 \cite{azcTow2}.
However it is unclear whether these properties are common in curved spaces;
any $p$-form charges (except scalars) can not be centers
in these backgrounds,
and string charges in AdS space are given as trivial elements of CE cohomology
 \cite{Berk,RWS,HKS,HKAdS,WZAdS}.

There are several attempts to evaluate brane charges 
in superalgebras
 in the matrix theory approach for pp-wave backgrounds in 
earlier works \cite{Sugi} and in \cite{Hyun2} and
in the Green-Schwarz type action approach for the AdS space \cite{WZAdS},
where new contributions to brane charges are obtained
in addition to the flat part.
It is desirable to have  background dependence of a brane charge,
since a brane charge is proportinal to a brane volume 
which depends on a background. 
However these brane charges are not allowed by
the central extension of the superalgebra.
Possible extension of the super-AdS algebra is also discussed \cite{KSosp}.
In order to clarify background dependence of a brane charge
and consistency of superalgebras,
we analyze a simple case  and show how these issues incorporate.
We use the supermatrix-valued coordinates
introduced in ``supertwistor formulation" \cite{RWS}
and compute brane charges analogously to \cite{WZAdS}.
We take a D0-brane case in the AdS$_2\times$S$^2$ \cite{Zhou}
which is the near horizon geometry of the Reissner-Nordstr$\ddot{\rm o}$m
black hole in four dimensions.
Its pp-wave limit and a flat limit are also examined. 

The organization of this paper is as follows.
In section 2.1, we explain the coset construction 
of the Wick rotated AdS$_2\times$S$^2$ space
 which is expressed by a coset 
$GL(2{\mid}2)/(GL(1))^4$ \cite{RWS}.    
Its Penrose limit \cite{HKSpp} is taken for
the bosonic part of  AdS$_2\times$S$^2$ 
in section 2.2
and for whole part in section 2.3.
In section 3 an action for a D0-brane in the
AdS$_2\times$S$^2$ space is given. 
The pseudo-supersymmetric WZ term is 
confirmed to satisfy three criteria;
1. its exterior derivative
to be correct two-form field strength, 
2. kappa-invariance of the action,
3. correct flat limit. 
Its pp-wave limit and a flat limit
are also given.
In section 4 supercharges are obtained and 
brane charges are calculated by anticommutators 
of these supercharges.
The BPS condition and the BPS mass are also discussed.

\section{ Super-AdS$_2\times$S$^2$ and its Penrose limit  }\par

Branes in the super-AdS$_2\times$S$^2$ background is expressed by the
coset 
\bea
SU(1,1{\mid}2)/SO(1,1)\times SO(2)~\longrightarrow~GL(2{\mid}2)/(GL(1))^4
\eea 
where  Wick rotations are performed and scaling degrees of freedom are introduced.
An element of the coset is denoted by $Z_M{}^A$ with indices $M$ for 
$GL(2{\mid}2)$ and indices $A$ for $(GL(1))^4$.
Two $GL(1)$'s are Lorentz groups of 
two 2-dimensional pseudospheres and other two $GL(1)$'s
are dilatations. 
At first we examine the bosonic part
of the AdS$_2\times$S$^2$ space,
$[GL(2)/(GL(1))^2]^2$, in section 2.1  
and take the Penrose limit
to derive the pp-wave metric in section 2.2.
The Penrose limit of the whole supergroup 
is presented in section 2.3.

\subsection{ AdS$_2\times$S$^2$}

 The bosonic part
of the AdS$_2\times$S$^2$ space is
$[GL(2)/GL(1)^2]^2$.
Coset elements  $z_m{}^a$ and $z_{\bar{m}}{}^{\bar{a}}$
correspond to a coset $GL(2)/(GL(1))^2$ 
for AdS$_2$ and another one for
S$^2$ respectively.
The left-invariant (LI) one forms are given by
\bea
j_a{}^b=z_a{}^mdz_m{}^b~~,~~
j_{\bar{a}}{}^{\bar{b}}=z_{\bar{a}}{}^{\bar{m}}dz_{\bar{m}}{}^{\bar{b}}
\label{jjbar}
\eea
where $z_a{}^m$ and $z_{\bar{a}}{}^{\bar{m}}$ are inverse of 
$z_m{}^a$ and $z_{\bar{m}}{}^{\bar{a}}$.
These indices are contracted with the Lorentz invariant metric
$\delta^{ab}$ and $\delta^{\bar{a}\bar{b}}$
and the Lorentz $GL(1)$ is 
assigned to $SO(2)$.
A LI current $j_{ab}$ is decomposed into three parts;
a trace part $\delta_{ab}j$,
a traceless symmetric part $j_{\langle ab \rangle}$
and an antisymmetric part $j_{[ab]}$,
\bea
j_{ab}
=\frac{1}{2}\delta_{ab}j+j_{\langle ab\rangle}
+j_{[ab]}~~~.
\eea  
If the following parameterization is chosen
\bea
z_m{}^a=R\delta_m^{a}+\left(
\begin{array}{cc}
x^0&x^1\\x^1&-x^0
\end{array}
\right)\label{bosonz}
\eea
whose inverse is
\bea
z_a{}^m=\frac{1}{R^2-(x^0)^2-(x^1)^2}[
R\delta_a^m{}-\left(
\begin{array}{cc}
x^0&x^1\\x^1&-x^0
\end{array}
\right)]~~~,
\eea
then the LI one forms are given by
\bea
j_{\langle ab\rangle}
=\frac{R}{R^2-\sum_\mu(x^\mu)^2}\left(
\begin{array}{cc}
dx^0&dx^1\\dx^1&-dx^0
\end{array}
\right)~~~\label{bosonj}.
\eea
The metric of this Wick rotated AdS$_2\times$S$^2$ is given by
\bea
ds^2_{\rm ``{\rm AdS}_2\times{\rm S}^2"}&=&
-\frac{1}{2}(
j_{\langle ab\rangle}j^{\langle ab\rangle}-j_{\langle \bar{a}\bar{b}\rangle}j^{\langle \bar{a}\bar{b}\rangle})
\nn\\
&=&-\frac{R^2}{[R^2-\sum_\mu(x^\mu)^2]^2}\displaystyle\sum_{\mu=0,1} (dx^\mu)^2
+\frac{R^2}{[R^2-\sum_\mu(x^\mu)^2]^2}\displaystyle\sum_{\mu=2,3} (dx^\mu)^2
\label{dsads}
\eea
expressing two 2-dimensional pseudospheres with the same unit curvature,
where the supersymmetry is imposed to restrict  curvatures to be same.
\vskip 6mm  
  
\subsection{ Penrose limit of AdS$_2\times$S$^2$}  

Under the global $GL(2)$ which is generated by $G_m{}^n$'s
\bea
G_m{}^n=z_m{}^a\pi_a{}^n ~~,~~
\left[z_m{}^a,\pi_b{}^n\right]=\delta_b^a\delta_m^n~~,~~
\left[G_m{}^n,G_l{}^k
\right]=-\delta_l^nG_m{}^k+\delta_m^kG_l{}^n
\eea
with the conjugate momenta $\pi$'s,
a contravariant vector 
\bea
w_{\langle mn\rangle}=z_{\langle m}{}^az_{n \rangle}{}_a\label{wzz}
\eea
is transformed as
\bea
\delta_\lambda  w_{mn}=\left[w_{ mn},\lambda_k{}^lG_l{}^k\right]
=\lambda_{m}{}^l w_{ln}
+\lambda_{n}{}^l w_{lm}
~~~.
\eea 
If we parameterize $z_m{}^a$ as \bref{bosonz},
$w_{\langle mn \rangle}$ becomes
\bea
w_{\langle mn\rangle}=
2R\left(
\begin{array}{cc}
x^0&x^1\\x^1&-x^0
\end{array}
\right)~~~\label{bosonw}~~~.
\eea 
So the $GL(2)$ generators are parameterized as
\bea
G^{mn}=
\left(
\begin{array}{cc}
D+P_0&P_1+M_{01}\\P_1-M_{01}&D-P_0
\end{array}
\right)~~~\label{bosonG}~~~,
\eea
where $D,P,M$ are dilatation, translations, Lorentz generators
respectively. The same parameterization is chosen 
for another $GL(2)$ generators $G^{\bar{m}\bar{n}}$
with replacing $P_0,P_1,M_{01}$ by $P_3,P_2,M_{32}$.

The Penrose limit is taken as follows:
Introduce the lightcone variables and scale them with the
parameter $\Omega$ then take a limit $\Omega\to 0$
\bea
&&P_\pm=\frac{1}{\sqrt{2}}(P_3\pm P_0)
\label{ppomega}\\
&&P_+\to\Omega^{-2}P_+~~,~~
(P_i,~M_{01},~M_{32})\to \Omega^{-1}(P_i,~M_{01},~M_{32})~_{i=1,2}~,~~
(P_-,~D)\to (P_-,~D)\nn
\eea
and therefore
\bea
x^{\pm}=\frac{1}{\sqrt{2}}(x^3\pm x^0)~~,~~
x^+\to \Omega^2 x^+~,~x^i\to \Omega x^i~,~
x^-\to x^-~~.\label{pomg}
\eea
The metric \bref{dsads}, which is scaled as $ds^2\to\Omega^{2}{ds'}^2$ and omit 
prime from now on,
 becomes  
\bea
ds^2_{\rm pp}&=&\Omega^{-2}(-\frac{1}{2})[
(j^{\langle ab\rangle})^2-(j^{\langle \bar{a}\bar{b}\rangle})^2]\left|_{\rm rescaling~\bref{pomg},~\Omega\to 0}\right.
\nn\\
&=&\displaystyle\frac{R^2}{[R^2-\frac{1}{2}{x^-}^2]^2}
\left\{
2dx^+dx^-+\displaystyle\sum_{i=1,2}(dx^i)^2+
\displaystyle
\frac{2dx^+dx^-+\sum_{i=1,2}(x^i)^2}
{R^2-\frac{1}{2}\left({x^-}\right)^2}
(dx^-)^2
\right\}\nn\\\label{spp1}
\eea
with $x^i=(ix^1,x^2)$. 
After field redefinition
\bea
\left\{\begin{array}{ccl}
X^i&=&\displaystyle\frac{R}{R^2-\frac{1}{2}\left({x^-}\right)^2}x^i\\
X^+&=&
\displaystyle\frac{R}{R^2-\frac{1}{2}\left({x^-}\right)^2}
\left(x^+
-\frac{1}{2}
\displaystyle\frac{x^-}{R^2-\frac{1}{2}\left({x^-}\right)^2}(x^i)^2
\right)\\
X^-&=&\frac{1}{2}\ln \displaystyle\frac{\sqrt{2}R+x^-}{\sqrt{2}R-x^-}
\end{array}\right.~~,\label{ppxx}
\eea
\bref{spp1} recast into the familiar form
\bea
ds^2_{\rm pp}&=&2dX^+X^-+\displaystyle\sum_{i=1,2}(dX^i)^2
+\frac{2}{R^2}\displaystyle\sum_{i=1,2}(X^i)^2(dX^-)^2~~.
\label{spp2}
\eea

A flat limit is taken by rescaling $x^\mu \to (1/R)x^\mu$ and
 $ds^2\to (1/R^2)ds^2$ then $R\to \infty$.
The metric \bref{dsads} has already dilatation scale $R$ exists
which is identified to the raduis of the curvature,
\bea
ds^2_{\rm flat}&=&
R^2\left(-\frac{1}{2}\right)
[j_{\langle ab\rangle}j^{\langle ab\rangle}
-j_{\langle \bar{a}\bar{b}\rangle}j^{\langle \bar{a}\bar{b}\rangle}]\left.{}\right|_{R\to \infty}
\nn\\
&=&\left\{-\frac{1}{[1-\sum_\mu(x^\mu/R)^2]^2}\displaystyle\sum_{\mu=0,1} (dx^\mu)^2
+\frac{1}{[1-\sum_\mu(x^\mu/R)^2]^2}\displaystyle\sum_{\mu=2,3} (dx^\mu)^2\right\}
\left.{}\right|_{R\to \infty}\nn\\
&=&\displaystyle\sum_{\mu=0,1,2,3}(dx^\mu)^2~~.
\label{dsflat}
\eea

\vskip 6mm
\subsection{ Penrose limit of super-AdS$_2\times$S$^2$}

Now we consider a coset $GL(2{\mid}2)/(GL(1))^4$
whose element is denoted by $Z_M{}^A$,
 and the supergroup $GL(2{\mid}2)$ is generated by
\bea
&&G_M{}^N=Z_N{}^A\Pi_A{}^N ~~,~~\left[Z_M{}^A,\Pi_B{}^N\right\}
=(-1)^A\delta_B^A\delta_M^N~\nn\\
&&\left[G_M{}^N,G_L{}^K
\right\}=(-1)^N(-\delta_L^NG_M{}^K+\delta_M^KG_L{}^N)~~~
\eea
where conjugate momenta $\Pi$'s are introduced.
Corresponding to that bosonic lightcone generators are defined as
\bref{ppomega} for generators of $GL(2)$  \bref{bosonG}, 
fermionic lightcone generators are identified from
the above algebra;
for example, $\left\{G_{\bar{1}}{}^{{1}},G_{1}{}^{\bar{1}}
\right\}\propto P_-~\Rightarrow$  $~
G_1{}^{\bar{1}},G_{\bar{1}}{}^1~\in Q_-
$,
$\left\{G_{\bar{2}}{}^1,G_1{}^{\bar{2}}
\right\}\propto P_+~\Rightarrow$ $
G_1{}^{\bar{2}},G_{\bar{2}}{}^1~\in Q_+
$ and so on.
A center $(D-\bar{D})={\rm Str} G_{MN}$ can be neglected.
Whole identification of 
lightcone generators with
 $GL(2{\mid}2)$ is given as follows:
\begin{center}
$G_{MN}=$
\begin{tabular}{
>{\columncolor[rgb]{1,.8,.5}}c >{\columncolor[rgb]{1,.8,.5}}c >{\columncolor[rgb]{1,.8,.5}}c
>{\columncolor[rgb]{1,.8,.5}}c
}
$D+P_0$& 
\multicolumn{1}{>{\columncolor[rgb]{.8,1,1}}c}{ $P_1+M_{01}$ } 
& \multicolumn{1}{>{\columncolor[rgb]{1,1,.5}}c}{ $Q_-$ } &
\multicolumn{1}{>{\columncolor[rgb]{.8,1,1}}c}{ $Q_+$ } 
\\
\multicolumn{1}{>{\columncolor[rgb]{.8,1,1}}c}{ $P_1-M_{01}$ } 
&$D-P_0$&
\multicolumn{1}{>{\columncolor[rgb]{.8,1,1}}c}{ $Q_+$ } &
\multicolumn{1}{>{\columncolor[rgb]{1,1,.5}}c}{ $Q_-$ }\\
\multicolumn{1}{>{\columncolor[rgb]{1,1,.5}}c}{ $Q_-$ }&
\multicolumn{1}{>{\columncolor[rgb]{.8,1,1}}c}{ $Q_+$ }&
$\bar{D}+P_3$&
\multicolumn{1}{>{\columncolor[rgb]{.8,1,1}}c}{ $P_2+M_{32}$ } \\
\multicolumn{1}{>{\columncolor[rgb]{.8,1,1}}c}{ $Q_+$ } &
\multicolumn{1}{>{\columncolor[rgb]{1,1,.5}}c}{ $Q_-$ }&
\multicolumn{1}{>{\columncolor[rgb]{.8,1,1}}c}{ $P_2-M_{32}$ } &
$\bar{D}-P_3$
\end{tabular}
\end{center}
\vskip 4mm
\noindent
where $P_0=(P_+-P_-)/\sqrt{2}$ and
$P_3=(P_++P_-)/\sqrt{2}$. 
The Penrose limit is given by rescaling
\bref{ppomega} and 
\bea
Q_+\to\Omega^{-1} Q_+~~,~~Q_-\to Q_-~~~,
\eea
or equivalently rescaling of
 $GL(2{\mid}2)$ generators 
as follows 
\bea
\begin{array}{c}
G_{MN}\to
\Omega^{-2}
\begin{tabular}{
>{\columncolor[rgb]{1,.8,1}}c >{\columncolor[rgb]{1,.8,1}}c >{\columncolor[rgb]{1,.8,1}}c
>{\columncolor[rgb]{1,.8,1}}c
}
$P_+$& 
\multicolumn{1}{>{\columncolor[rgb]{.8,1,1}}c}{ } 
& \multicolumn{1}{>{\columncolor[rgb]{1,1,.5}}c}{ } &
\multicolumn{1}{>{\columncolor[rgb]{.8,1,1}}c}{  } 
\\
\multicolumn{1}{>{\columncolor[rgb]{.8,1,1}}c}{  } 
&$P_+$&
\multicolumn{1}{>{\columncolor[rgb]{.8,1,1}}c}{ } &
\multicolumn{1}{>{\columncolor[rgb]{1,1,.5}}c}{  }\\
\multicolumn{1}{>{\columncolor[rgb]{1,1,.5}}c}{  }&
\multicolumn{1}{>{\columncolor[rgb]{.8,1,1}}c}{  }&
$P_+$&
\multicolumn{1}{>{\columncolor[rgb]{.8,1,1}}c}{ } \\
\multicolumn{1}{>{\columncolor[rgb]{.8,1,1}}c}{  } &
\multicolumn{1}{>{\columncolor[rgb]{1,1,.5}}c}{  }&
\multicolumn{1}{>{\columncolor[rgb]{.8,1,1}}c}{  } &
$P_+$
\end{tabular}
+\Omega^{-1}
\begin{tabular}{
>{\columncolor[rgb]{1,.8,.5}}c >{\columncolor[rgb]{1,.8,.5}}c >{\columncolor[rgb]{1,.8,.5}}c
>{\columncolor[rgb]{1,.8,.5}}c
}
& 
\multicolumn{1}{>{\columncolor[rgb]{.8,1,1}}c}{$P_1,M$ } 
& \multicolumn{1}{>{\columncolor[rgb]{1,1,.5}}c}{ } &
\multicolumn{1}{>{\columncolor[rgb]{.8,1,1}}c}{$Q_+$  } 
\\
\multicolumn{1}{>{\columncolor[rgb]{.8,1,1}}c}{ $P_1,M$ } 
&&

\multicolumn{1}{>{\columncolor[rgb]{.8,1,1}}c}{ $Q_+$ } &
\multicolumn{1}{>{\columncolor[rgb]{1,1,.5}}c}{ }\\
\multicolumn{1}{>{\columncolor[rgb]{1,1,.5}}c}{  }&
\multicolumn{1}{>{\columncolor[rgb]{.8,1,1}}c}{$Q_+$   }&
&
\multicolumn{1}{>{\columncolor[rgb]{.8,1,1}}c}{$P_2,\bar{M}$ } \\
\multicolumn{1}{>{\columncolor[rgb]{.8,1,1}}c}{ $Q_+$  } &
\multicolumn{1}{>{\columncolor[rgb]{1,1,.5}}c}{  }&
\multicolumn{1}{>{\columncolor[rgb]{.8,1,1}}c}{ $P_2,\bar{M}$ } &
\end{tabular}
\\\\
+\Omega^{0}~~
\begin{tabular}{
>{\columncolor[rgb]{1,1,.5}}c >{\columncolor[rgb]{1,1,.5}}c >{\columncolor[rgb]{1,1,.5}}c
>{\columncolor[rgb]{1,1,.5}}c
}
$D,P_-$& 
\multicolumn{1}{>{\columncolor[rgb]{.8,1,1}}c}{ } 
& \multicolumn{1}{>{\columncolor[rgb]{1,1,.5}}c}{$ Q_-$} &
\multicolumn{1}{>{\columncolor[rgb]{.8,1,1}}c}{  } 
\\
\multicolumn{1}{>{\columncolor[rgb]{.8,1,1}}c}{  } 
&$D,P_-$&
\multicolumn{1}{>{\columncolor[rgb]{.8,1,1}}c}{ } &
\multicolumn{1}{>{\columncolor[rgb]{1,1,.5}}c}{ $ Q_-$ }\\
\multicolumn{1}{>{\columncolor[rgb]{1,1,.5}}c}{ $ Q_-$ }&
\multicolumn{1}{>{\columncolor[rgb]{.8,1,1}}c}{  }&
$\bar{D},P_-$&
\multicolumn{1}{>{\columncolor[rgb]{.8,1,1}}c}{ } \\
\multicolumn{1}{>{\columncolor[rgb]{.8,1,1}}c}{  } &
\multicolumn{1}{>{\columncolor[rgb]{1,1,.5}}c}{ $ Q_-$ }&
\multicolumn{1}{>{\columncolor[rgb]{.8,1,1}}c}{  } &
$\bar{D},P_-$
\end{tabular}
\end{array}
\label{pptbl}
\eea
and taking the limit $\Omega\to 0$.
\section{ Super-D0-brane actions}\par

The action of a super-D0-brane in AdS$_2\times$S$^2$ is given in \cite{Zhou}
and we reformulate this into the ``supertwistor formulation"
in this section.
We must construct the WZ term with taking care of the boundary term 
to obtain brane charges.
In section 3.1 we present a general argument
of the pseudo-supersymmetric WZ term \cite{WZAdS}.
Then the super-D0-brane action is examined explicitly in section 3.2.
\vskip 6mm
\subsection{ WZ term for a $p$-brane}

In order to compute a brane charge by the anticommutator of 
supercharges we need to construct the pseudo-supersymmetric WZ term. 
For an group element, $Z$,
it can be parameterized generally as
$
Z={\bf 1}+X
$
whose inverse is 
$
Z^{-1}=\displaystyle\sum_{n=0}^{\infty}(-X)^n
$.
A left-invariant (LI) one form current is given as
\bea
J=Z^{-1}dZ=\displaystyle\sum_{n=0}^{\infty}(-X)^n~dX\label{intro1}~~~,
\eea
satisfying the Maurer-Cartan equation,
$
dJ=-JJ~~\label{IntrMC}
$. 
The right hand side of the integration of the Maurer-Cartan equation is calculated as
\bea
-\int JJ=\int d\left(\displaystyle\sum_{n=1}^{\infty}(-X)^n~dX\right)
=\displaystyle\sum_{n=1}^{\infty}(-X)^n~dX+{\rm boundary~term}~~~.
\label{intro2}~~~
\eea
Its left-invariance requires
``${\rm boundary~term}=dX$", so 
``$-\displaystyle\int_{L.I.}JJ=J$".
On the other hand  the ``pseudo"-left-invariant one form is
\bea
-\int_{P.L.I.} JJ=J-dX~~~.
\eea

For super-$p$-branes the Wess-Zumino (WZ) term is given as
the one-dimensional higher space integration of 
LI $(p+2)$ form currents  
\bea
{\cal L}_{WZ}=B_{[p+1]}=\int_{P.L.I.} H_{[p+2]}~~,~~H_{[p+2]}=J^{p+2}~~~,
\eea
which is pseudo-supersymmetry invariant producing 
the topological $p$-brane charge.
This pseudo-invariance corresponds to
the choice of the boundary term of the integration to be zero.
If a LI $(p+1)$ form current exists
$
\tilde{B}_{[p+1]}=J^{p+1}~~
$ with $d\tilde{B}_{[p+1]}=H_{[p+2]}$,  
the pseudo-invariant $(p+1)$-form is given by \cite{WZAdS}
\bea
{\cal L}_{WZ}=B_{[p+1]}=\tilde{B}_{[p+1]}-d\left(X(dX)^p\right)~~.\label{LWZp}
\eea

The super-AdS group and the nondegenerate supertranslation group
are the cases where $\tilde{B}_{2}=(J)^2$ exists \cite{Berk,RWS,HKS,HKAdS,WZAdS}. 
For the nondegenerate supertranslation group
there exists the LI 2-form current
\bea
\tilde{B}_{[2]}=J_{\tilde{Q}}J_{Q}~~,~~d\tilde{B}_{[2]}=H_{[3]}=J_QJ_PJ_Q
\eea
where $\tilde{Q}$ is introduced to make a nondegenerate
fermionic group metric  \cite{Green} and satisfies
$
dJ_{\tilde{Q}}=J_{P}J_Q$.
The WZ term for a superstring in a flat space is given by
\bea
{\cal L}_{WZ}=B_{[2]}=\tilde{B}_{[2]}-d\left(X_{\tilde{Q}}dX_{Q}\right)~~~.
\eea
The auxiliary variable $X_{\tilde{Q}}$ is not 
contained in the kinetic term, 
so this can be gauged away in this case.

Now let us go back to the curved background cases. 
The supercoset construction is obtained by 
replacing bosonic currents \bref{jjbar}
with supercurrents 
\bea
J_A{}^B=Z_A{}^MdZ_M{}^B\label{superJ}~~,~~Z_M{}^A=
\left(
\begin{array}{cc}
{\bf 1}&\\&{\bf 1}
\end{array}
\right)
+
\left(
\begin{array}{cc}
X&\Theta\\\bar{\Theta}&\bar{X}
\end{array}
\right)~~~.
\eea
All components are contained in $X$ and $\bar{X}$ \cite{RWS,HKAdS},
especially the antisymmetric part
plays a role of auxiliary variables. 
The action for a D$p$-brane is given by
\bea
I_{{\rm D}p}&=&T\int~d^{p+1}\sigma ({\cal L}_{DBI}+{\cal L}_{WZ})
\eea 
where the Dirac-Born-Infeld term is given as
\bea
{\cal L}_{DBI}=-\sqrt{\det (g_{rs}+{\cal F}_{rs})}~~,~~
g_{rs}=(J_r^{\langle ab\rangle}J_{\langle ab\rangle,s}
-J_r^{\langle \bar{a}\bar{b}\rangle}J_{\langle \bar{a}\bar{b}\rangle,s})/{\rm tr}{\bf 1}
~~
\eea
with
$J=d\sigma^rJ_r$~,~${r=0,1,\cdots,~p+1}$,
the tension $T$ and the DBI field strength ${\cal F}_{rs}$.
The WZ term in super-AdS spaces has the form of \bref{LWZp} and it
 must satisfy following three criteria;
\begin{enumerate}
\item{Its exterior derivative produces the 
correct field strength, $d{\cal L}_{WZ,[p+1]}=H_{[p+2]}$ with $dH_{[p+2]}=0$.}
\item{The action is kappa-symmetric.}
\item{Its flat limit reduces into  the correct flat WZ term.}
\end{enumerate}

\vskip 6mm
\subsection{ Super-D0-brane action}

The action for a D0 brane is given as
\bea
I_{\rm D0}&=&\int d\tau~({\cal L}_{DBI,{\rm D0}}+{\cal L}_{WZ,{\rm D0}})\nn\\
{\cal L}_{DBI,{\rm D0}}&=&T \sqrt{g_{00}}
~~,~~g_{00}=(J_0^{\langle ab\rangle}J_{\langle ab\rangle,0}
-J_0^{\langle \bar{a}\bar{b}\rangle}J_{\langle \bar{a}\bar{b}\rangle,0})/2
\label{D0act}\\
{\cal L}_{WZ,{\rm D0}}&=&\left\{K\epsilon_{ab}\left(J_0^{ab}-d(\delta^{am}X_m{}^b)\right)
+\bar{K}\epsilon_{\bar{a}\bar{b}}
\left(J_0^{\bar{a}\bar{b}}-d(\delta^{\bar{a}\bar{m}}\bar{X}_{\bar{m}}{}^{\bar{b}})
\right)
\right\}
\nn~~~
\eea
where $K$ and $\bar{K}$ are determined by the kappa-symmetry as
\bea
T^2=K^2+\bar{K}^2~~~\label{Tkkbar}
\eea
which agrees with Zhou's result \cite{Zhou}.

This WZ term satisfies three criteria:
The criterion 1. is confirmed easier.
 The two-form RR-field strength for a D0-brane is given by
\bea
H_{[2]}=d{\cal L}_{WZ}=-\left\{K\epsilon_{ab}
(J^{ac}J_c{}^b+J^{a\bar{c}}J_{\bar{c}}{}^b)
+\bar{K}\epsilon_{\bar{a}\bar{b}}
(J^{\bar{a}c}J_c{}^{\bar{b}}+J^{\bar{a}\bar{c}}J_{\bar{c}}{}^{\bar{b}})\right\}
\eea
where index $0$ of $J_{AB,0}$ has been omitted. 
The field strength $H_{[2]}$ in a flat space 
contains  bilinear of fermionic indices currents
only. On the other hand the one in AdS$^2\times$S$_{2}$ 
space contains bilinear of not only the fermionic indices currents 
but also the bosonic indices currents in order to make closed form $dH_{[2]}=0$
\cite{Zhou}.
For the expression of ${\cal L}_{WZ}$ in \bref{D0act} 
this closure is trivial, i.e. $dH_{[2]}=dd{\cal L}_{WZ}=0=ddJ$.

The criterion 2. is calculated as follows:
If we introduce the following combination for 
an arbitrary variation $\delta Z_M{}^A$
\bea
\Delta J^{AB}=Z^{AM}\delta Z_M{}^B~~,
\eea
the kappa-variation is characterized by vanishing bosonic components
\bea
\Delta_{{\kappa}}J^{ab}=0=\Delta_{{\kappa}}J^{\bar{a}\bar{b}}~~~
\eea
as usual. In particular in this parameterization
$\Delta_{{\kappa}}J^{a\bar{b}}={\kappa}^{a\bar{b}}$ and
$\Delta_{{\kappa}}J^{\bar{a}{b}}={\kappa}^{\bar{a}b}$ 
\footnote{
In the canonical computation the fermionic constraints 
are given as \cite{HKAdS}
\bea
\tilde{D}_{\bar{a}b}=D_{\bar{a}b}+\displaystyle\frac{\partial {\cal L}_{WZ}}{\partial \dot{\Theta}_m{}^{\bar{a}}}
X_m{}^b=0~~,~~
\tilde{D}_{a\bar{b}}=D_{a\bar{b}}+\displaystyle
\frac{\partial {\cal L}_{WZ}}{\partial \dot{\bar{\Theta}}_{\bar{m}}{}^a}
\bar{X}_{\bar{m}}{}^{\bar{b}}=0~~
\eea
with the local $GL(2{\mid}2)$ generators, $D_{AB}$'s.
The kappa-variation is 
generated by these fermionic constraints 
\bea
\delta_{\kappa} {\cal O}
=\left[{\cal O},F_{\bar{a}b}{\kappa}^{b\bar{a}}+F_{a\bar{b}}{\kappa}^{\bar{b}a}
\right]
~~~,~~~
\delta_{\kappa} Z_M{}^b=Z_M{}^{\bar{a}} {\kappa}_{\bar{a}}{}^b~~,~~
\delta_{\kappa} Z_M{}^{\bar{b}}=Z_M{}^{a} {\kappa}_{a}{}^{\bar{b}}~~
\eea
with these parameters are projected into half.
}.
Using the kappa-variation of currents, 
\bea
\delta_{{\kappa}}J^{ab}&=&J^{a\bar{c}}\Delta_{\kappa} J_{\bar{c}}{}^b
-(\Delta_{\kappa} J^{a\bar{c}})J_{\bar{c}}{}^b\label{Jkappa}
\eea
and the similar for the $J^{\bar{a}\bar{b}}$, 
the kappa-variation of the D0-brane action \bref{D0act} 
becomes
\bea
\delta_{\kappa}{\cal L}_{DBI}&=&\frac{T}{\sqrt{g_{00}}}
\left\{
\Delta_{\kappa} J_{a\bar{b}}~
\left(J^{ac}J^{\bar{b}}{}_c-J^{\bar{c}\bar{b}}J_{\bar{c}}{}^a\right)
+\Delta_{\kappa} J_{\bar{a}{b}}~
\left(J^{cb}J_c{}^{\bar{a}}-J^{\bar{a}\bar{b}}J^b{}_{\bar{b}}\right)
\right\}\nn\\
\label{kappaL0}\\
\delta_{\kappa}{\cal L}_{WZ}&=&-
\Delta_{\kappa} J_{a\bar{b}}~
\left(K\epsilon^{ab}J^{\bar{b}}{}_b+\bar{K}\epsilon^{\bar{a}\bar{b}}J_{\bar{a}}{}^a\right)
-\Delta_{\kappa} J_{\bar{a}{b}}~
\left(K\epsilon^{ab}J_a{}^{\bar{a}}+\bar{K}\epsilon^{\bar{a}\bar{b}}J^b{}_{\bar{b}}\right)
\label{kappaLWZ}\nn~~~.
\eea
The kappa-invariance of the action leads to
\bea
\delta_{\kappa}({\cal L}_{DBI}+{\cal L}_{WZ})&=&
(\Delta_{\kappa} J_{a\bar{b}}J_{\bar{c}d}
+\Delta_{\kappa} J_{\bar{c}d}J_{a\bar{b}})
{\cal P}^{a\bar{b}\bar{c}d}\nn\\
{\cal P}^{a\bar{b}\bar{c}d}&=&
\frac{T}{\sqrt{g_{00}}}(J^{ad}\delta^{\bar{b}\bar{c}}-
\delta^{ad}J^{\bar{c}\bar{b}})
-(K\epsilon^{ad}\delta^{\bar{b}\bar{c}}
+\bar{K}\delta^{ad}\epsilon^{\bar{c}\bar{b}})\nn\\
0&=&{\cal P}^{a\bar{b}\bar{c}d}\Delta_{\kappa} J_{a\bar{b}}=
{\cal P}^{a\bar{b}\bar{c}d}\Delta_{\kappa} J_{\bar{c}d}
\label{pkappa}~~.
\eea
Equivalently projected kappa-parameters
${\cal P}^{a\bar{b}\bar{c}d}{\kappa}_{a\bar{b}}=0=
{\cal P}^{a\bar{b}\bar{c}d}{\kappa}_{\bar{c}d}
$ with the relation \bref{Tkkbar}.

The criterion 3. is evaluated as follows:
Under the flat limit which is obtained by rescaling, 
$X,\bar{X}\to X/R,\bar{X}/R$, $\Theta, \bar{\Theta}\to \Theta/\sqrt{R},\bar{\Theta}/\sqrt{R}$ and
${\cal L}_{WZ}\to{\cal L}_{WZ}/R$ as well as ${\cal L}_{DBI,{\rm D0}}\to{\cal L}_{DBI,{\rm D0}}/R $,
and  $R\to \infty$,
the WZ term \bref{D0act} reduces into
:
\bea
{\cal L}_{WZ,{\rm flat}}&=&-
K\left\{\epsilon_{ab}\bar{\Theta}^{\bar{m}a} d \bar{\Theta}_{\bar{m}}{}^b
+\epsilon_{\bar{a}\bar{b}}
{\Theta}^{{m}\bar{a}} d {\Theta}_{{m}}{}^{\bar{b}}\right\}\label{WZflatD0}
\eea
where $K=\bar{K}$ is imposed by requiring 4-dimensional
covariance and this is consistent with \cite{D0p,HKD01}.

The WZ term for a D0-brane in the pp-wave background is obtained by
taking the Penrose limit \cite{HKSpp} given in \bref{pptbl}.
The WZ term is scaled under the Penrose limit as
${\cal L}_{WZ}~\to~\Omega{\cal L}_{WZ}$
as well as the DBI term ${\cal L}_{DBI}~\to~\Omega{\cal L}_{DBI}$.
Since the antisymmetric components of currents and parameters
are scaled as
$(J_{[ab]},J_{[\bar{a}\bar{b}]})\to\Omega(J_{[ab]},J_{[\bar{a}\bar{b}]})$, 
$([X],[\bar{X}])\to\Omega([X],[\bar{X}])$, 
all terms in ${\cal L}_{WZ}$ are survived. 
The LI current after taking the Penrose limit, 
which is still complicated function of $x^-=x_+$ 
and $\theta^-$,
is denoted as $J_{{\rm pp},AB}$.
Then the WZ term is written as
\bea
{\cal L}_{WZ,{\rm pp}}
&=&\left\{K\epsilon_{ab}\left(J_{{\rm pp}}^{ab}-d(\delta^{am}X_m{}^b)\right)
+\bar{K}\epsilon_{\bar{a}\bar{b}}
\left(J_{{\rm pp}}^{\bar{a}\bar{b}}-d(\delta^{\bar{a}\bar{m}}\bar{X}_{\bar{m}}{}^{\bar{b}})
\right)
\right\}~~~.\nn\\
\label{ppwz}
\eea

\vskip 6mm
\section{ D0-brane charges}\par

The Noether charges of global $GL(2{\mid}2)$ symmetry with parameters $\Lambda^{MN}$
are modified under the existence of the WZ term \cite{azcTow}
\bea
\Lambda^{NM}G_{MN}~\to~\Lambda^{NM}\tilde{G}_{MN}=\Lambda^{NM}G_{MN}-U_\Lambda~~,~~
\delta_\Lambda{\cal L}=dU_\Lambda\label{WZ}~~~.
\eea 
Because of $\delta_\Lambda {\cal L}_{DBI}=0=\delta_\Lambda J$,
 only the total derivative terms in the WZ term in \bref{D0act} contribute to $U$
which is given as
\bea
U_{\Lambda}&=&-\left(
K\epsilon_{ab}\delta^{am}\delta_\Lambda X_m{}^b
+\bar{K}\epsilon_{\bar{a}\bar{b}}\delta^{\bar{a}\bar{m}}\delta_\Lambda \bar{X}_{\bar{m}}{}^{\bar{b}}
\right)~~,\\
\delta_\Lambda Z_M{}^A&=&\Lambda_M{}^L Z_L{}^A~~~.
\eea
This modified $GL(2{\mid}2)$ generators are then expressed as
\bea
&&\tilde{G}_{MN}=G_{MN}+{\cal T}_{MN}\nn\\
&&\left\{\begin{array}{lcl}
{\cal T}_{mn}=-K(\delta_m^a+X_m{}^a)\epsilon_{ab}\delta^b_n~~&,&
{\cal T}_{m\bar{n}}=-\bar{K}\Theta_m{}^{\bar{a}}\epsilon_{\bar{a}\bar{b}}\delta^{\bar{b}}_{\bar{n}}~\\
{\cal T}_{\bar{m}\bar{n}}=-\bar{K}
(\delta_{\bar{m}}^{\bar{a}}+\bar{X}_{\bar{m}}{}^{\bar{a}})
\epsilon_{\bar{a}\bar{b}}\delta^{\bar{b}}_{\bar{n}}~~&,&
{\cal T}_{\bar{m}n}=-K\bar{\Theta}_{\bar{m}}{}^a
\epsilon_{ab}\delta^b_n
\end{array}\right.~~~\label{calT}~,
\eea
and also re-expressed as
\bea
&&\tilde{G}_{MN}=Z_M{}^A\tilde{\Pi}_{AN}\nn\\
&&\left\{\begin{array}{lcl}
\tilde{\Pi}_{an}=\Pi_{an}-K\epsilon_{ab}\delta^b_{n}~~&,&
\tilde{\Pi}_{a\bar{n}}=\Pi_{a\bar{n}}
\\
\tilde{\Pi}_{\bar{a}\bar{n}}=\Pi_{\bar{a}\bar{n}}-\bar{K}\epsilon_{\bar{a}\bar{b}}
\delta^{\bar{b}}_{\bar{n}}
~~&,&\tilde{\Pi}_{\bar{a}{n}}=\Pi_{\bar{a}{n}}
\end{array}\right.\label{tildepi}
\nn
\eea
satisfying the same algebra of $GL(2{\mid}2)$ \footnote{
We thank Kiyoshi Kamimura for pointing out that a total derivative term
does not change the global symmetry algebra.}.
Under the existence of the WZ term  Noether charges 
$\tilde{G}_{MN}$ are conserved and generate the closed superalgebra. 
Fermionic parts are identified as supercharges as
$\tilde{G}_{m\bar{n}}\equiv Q_{{\rm AdS},m\bar{n}}$,
 $\tilde{G}_{\bar{m}n}\equiv \bar{Q}_{{\rm AdS},\bar{m}n}$,

while bosonic parts are interpreted as sum of the 
bosonic charges $G_{MN}$ and 
topological charges ${\cal T}_{mn},~{\cal T}_{\bar{m}\bar{n}}$
conventionally.
Therefore commutators of supercharges for a D0-brane in the AdS$_2\times$S$^2$ background
are given by
\bea
\left\{Q_{\rm AdS},Q_{\rm AdS}\right\}&=&0=\left\{\bar{Q}_{\rm AdS},\bar{Q}_{\rm AdS}\right\}\nn\\
\left\{ \bar{Q}_{{\rm AdS},\bar{m}}{}^n, Q_{{\rm AdS},l}{}^{\bar{k}}
\right\}&=&\left\{ Q_{{\rm AdS},l}{}^{\bar{k}} , \bar{Q}_{{\rm AdS},\bar{m}}{}^n \right\}
~\label{SUSYads}\\&=&
-\delta_l^n
(P_{\bar{m}}{}^{\bar{k}}+{M}_{\bar{m}}{}^{\bar{k}}+{\cal T}_{\bar{m}}{}^{\bar{k}})
+\delta_{\bar{m}}^{\bar{k}}(P_l{}^n+M_l{}^n+{\cal T}_l{}^n)
+\delta_l^n\delta_{\bar{m}}^{\bar{k}}(\bar{D}-D)
~~\nn
\eea
where a center $\bar{D}-D={\rm Str} G_{MN}$ can be ignored.
The D0-brane charges are ${\cal T}_{mn}$ and ${\cal T}_{\bar{m}\bar{n}}$.
They have position dependence so they are not center.
The generators of this algebra are 
rather  $Q,\bar{Q},\tilde{G}_{mn},\tilde{G}_{\bar{m}\bar{n}}$
than $Q,\bar{Q},P,M,{\cal T}$.

Under the flat limit 
contributions from the WZ term are given as
\bea
\left\{\begin{array}{lcl}
{\cal T}_{\rm flat,}{}_{mn}=-K\delta_m^a\epsilon_{ab}\delta_n^b~~&,&
{\cal T}_{\rm flat,}{}_{m\bar{n}}=-\bar{K}\Theta_m^{\bar{a}}\epsilon_{\bar{a}\bar{b}}\delta_{\bar{n}}^{\bar{b}}
~~\\\\
{\cal T}_{\rm flat,}{}_{\bar{m}\bar{n}}=-\bar{K}\delta_{\bar{m}}^{\bar{a}}\epsilon_{\bar{a}\bar{b}}
\delta_{\bar{n}}^{\bar{b}}~~&,&
{\cal T}_{\rm flat,}{}_{\bar{m}n}=-K\bar{\Theta}_{\bar{m}}^a\epsilon_{ab}\delta_n^b
\end{array}\right.
~~\label{ttflat}
\eea
and imposing four-dimensional symmetry leads to $K=\bar{K}$. 
The commutators of supercharges \bref{SUSYads}
reduce into
as
\bea
\left\{Q_{\rm flat},Q_{\rm flat}\right\}&=&0=\left\{\bar{Q}_{\rm flat},\bar{Q}_{\rm flat}\right\}\nn\\
\left\{\bar{Q}_{\rm flat,}{}_{\bar{m}}{}^n,Q_{\rm flat,}{}_l{}^{\bar{k}}\right\}
&=&\left\{Q_{\rm flat,}{}_l{}^{\bar{k}},\bar{Q}_{\rm flat,}{}_{\bar{m}}{}^n\right\}\nn\\
&=&
-\delta_l^n(P_{\bar{m}}{}^{\bar{k}}+{\cal T}_{{\rm flat},\bar{m}}{}^{\bar{k}})
+\delta_{\bar{m}}^{\bar{k}}(P_l{}^n+{\cal T}_{{\rm flat},l}{}^n)~~~.
\label{SUSYflat}
\eea
This is the supertranslation algebra
with a center for the D0-brane.

Under the Penrose limit \bref{pptbl} 
contributions from the WZ term are given as
\bea
\left\{\begin{array}{ccl}
{\cal T}_{{\rm pp},mn}=-K({\bf 1}+X\left.{}\right|_{x^+=0})_m{}^a\epsilon_{ab}\delta^b_n\displaystyle&,&
{\cal T}_{{\rm pp},m\bar{n}}=-\bar{K}\Theta_m{}^{\bar{a}}\epsilon_{\bar{a}\bar{b}}\delta^{\bar{b}}_{\bar{n}}\\\\
{\cal T}_{{\rm pp},\bar{m}\bar{n}}=-\bar{K}
({\bf 1}+\bar{X}\left.{}\right|_{x^+=0})_{\bar{m}}{}^{\bar{a}}
\epsilon_{\bar{a}\bar{b}}\delta^{\bar{b}}_{\bar{n}}
\displaystyle&,&
{\cal T}_{{\rm pp},\bar{m}n}=-K\bar{\Theta}_{\bar{m}}{}^{a}\epsilon_{ab}\delta^b_n
\end{array}\right.~~~.\label{calTpp}
\eea
The commutators of supercharges
become
\bea
\left\{Q_{\rm pp},Q_{\rm pp}\right\}&=&0=\left\{\bar{Q}_{\rm pp},\bar{Q}_{\rm pp}\right\}\nn\\
\left\{\bar{Q}_{\rm pp,}{}_{\bar{m}}{}^n,Q_{\rm pp,}{}_l{}^{\bar{k}}\right\}
&=&\left\{Q_{\rm pp,}{}_l{}^{\bar{k}},\bar{Q}_{\rm pp,}{}_{\bar{m}}{}^n\right\}\nn\\
&=&
-\delta_l^n(G_{\rm pp,}{}_{\bar{m}}{}^{\bar{k}}+{\cal T}_{\rm pp,}{}_{\bar{m}}{}^{\bar{k}})
+\delta_{\bar{m}}^{\bar{k}}(G_{\rm pp,}{}_l{}^n+{\cal T}_{\rm pp,}{}_l{}^n)~~~,
\label{SUSYppG}
\eea
and the concrete expression is given by
\bea
\left\{\begin{array}{l}
\left\{\bar{Q}_{{\rm pp},+,}{}_{\bar{m}}{}^n,Q_{{\rm pp},+,}{}_l{}^{\bar{k}}\right\}
~~=~~-\delta_l^n
\left(
\begin{array}{cc}
\sqrt{2}P_+&\\&-\sqrt{2}P_+
\end{array}
\right){}_{\bar{m}}{}^{\bar{k}}\\\\
\left\{\bar{Q}_{{\rm pp},-,}{}_{\bar{m}}{}^n,Q_{{\rm pp},-,}{}_l{}^{\bar{k}}\right\}
~~=~~
-\delta_l^n
\left\{\left(
\begin{array}{cc}
\sqrt{2}P_-&\\&-\sqrt{2}P_-
\end{array}
\right)\right.
\\\\ 
~~~~~~~~~\left.
-\left(
\begin{array}{cc}
&Kx^1-\bar{K}x^2\\Kx^1-\bar{K}x^2&
\end{array}
\right)
\left(

\begin{array}{cc}
&1\\-1&
\end{array}
\right)
\right\}{}_{\bar{m}}{}^{\bar{k}}
+\delta^n_l\delta_{\bar{m}}^{\bar{k}}(D-\bar{D})
\\\\
\left\{\bar{Q}_{{\rm pp},+,}{}_{\bar{m}}{}^n,Q_{{\rm pp},-,}{}_l{}^{\bar{k}}\right\}\\\\
~~~~~~~
=
-\delta_l^n
 \left\{\left(
 \begin{array}{cc}
 &P_2+M_{32}\\P_2-M_{32}
 \end{array}
 \right)
 -\bar{K}\left(
 \begin{array}{cc}
 1+\frac{x^-}{\sqrt{2}}&\\&1-\frac{x^-}{\sqrt{2}}
 \end{array}
 \right)
 \left(
 \begin{array}{cc}
 &1\\-1&
 \end{array}
 \right)
 \right\}{}_{\bar{m}}{}^{\bar{k}}\\\\
 ~~~~~~~~~+\delta_{\bar{m}}{}^{\bar{k}}
 \left\{\left(
 \begin{array}{cc}
 &P_1+M_{01}\\P_1-M_{01}
 \end{array}
 \right)
-K\left(
 \begin{array}{cc}
 1-\frac{x^-}{\sqrt{2}}&\\&1+\frac{x^-}{\sqrt{2}}
 \end{array}
 \right)
 \left(
 \begin{array}{cc}
 &1\\-1&
 \end{array}
 \right)
 \right\}{}_l{}^n
 \end{array}\right.\label{SUSYpp}
\eea
The anticommutator 
$
\left\{\bar{Q}_{{\rm pp},-,}{}_{\bar{m}}{}^n,Q_{{\rm pp},+,}{}_l{}^{\bar{k}}\right\}$
satisfies the same relation with the last equation of \bref{SUSYpp}
with opposite lightcone projection.
This is the anticommutators of supercharges for a D0-brane 
in the 2+2 dimensional pp-wave background,
where the super-pp-algebra part is consistent with \cite{jose,HKSPen}
 by setting a center $D-\bar{D}={\rm Str} G_{MN}$ 
to be zero. 
The D0-brane charge in the pp-wave background is independent 
on $x^+$ but depend on $x^i$'s ($i=1,2$) and 
$x^-$ ($x^-=x_+=p_+\tau$ in the lightcone gauge).
In the lightcone gauge all
supersymmetries are broken except $x^i=0$,
since  $Q_+$ are broken because of $P_+\neq 0$
and $Q_-$  are also broken for $x^i\neq 0$ \cite{ske}.



In order to examine the BPS condition let us compare with
the case of a D0-brane in a flat space. 
The supercharge $Q_\alpha$ and the fermionic constraint $\tilde{D}_\alpha$ 
satisfy the following anticommutators \cite{HKD01}
\bea
&&\left\{Q_\alpha,Q_\beta\right\}=-2C(P_\mu\gamma^\mu-T\gamma^{11})
= -2M_{\rm BPS}{\cal P}_{+,\alpha\beta}\nn\\
&&~~~~~\Rightarrow \left\{\begin{array}{ccl}
Q_+=Q{\cal P}_+&\cdots&{\rm broken}~~\Rightarrow~~\theta^+~{\rm Nambu}\mbox{-}{\rm Goldstone~mode}\\
Q_-=Q{\cal P}_-&\cdots&{\rm unbroken}~~\Rightarrow~~\theta^-~{\rm independent ~observables}
\end{array}\right.\label{QQDD}\\
&&\left\{\tilde{D}_\alpha,\tilde{D}_\beta\right\}=2C(P_\mu\gamma^\mu-T\gamma^{11})= 
2M_{\rm BPS}{\cal P}_{+,\alpha\beta}\nn\\
&&~~~~~\Rightarrow\left\{\begin{array}{ccl}
\tilde{D}_+=\tilde{D}{\cal P}_+&\cdots&{\rm second ~class}~\Rightarrow~1/2~ ({\rm in~ phase~space})~\theta^+{\rm physical}\\
\tilde{D}_-=\tilde{D}{\cal P}_-&\cdots&{\rm first~class},~{\kappa}~{\rm symmetry}~\Rightarrow~
\theta^-~{\rm gauged ~away}
\end{array}\right.\nn
\eea
with $\left\{Q_\alpha,\tilde{D}_\beta\right\}=0$. 
The right hand side of $\left\{Q,Q\right\}$ is the BPS projection operator
while the one of $\left\{\tilde{D},\tilde{D}\right\}$ is the kappa-symmetry projection operator.
For a particle case they coincide, but in general they are 
  similar  but not equal \cite{HKD01}.
The BPS projection operator is ${\cal P}_{+,\alpha\beta}$
and the BPS mass is $M_{\rm BPS}\geq T$.

In flat space (mass)$^2$ classifies BPS states. 
In curved space a quadratic Casimir operator $c_{[2]}$,
which is the background  covariant Hamiltonian, 
corresponds to (mass)$^2$. 
The quadratic Casimir operator  in the super-AdS and the super-pp-wave
spaces are given as \cite{Me1,HKSpp}
$c_{[2],{\rm sAdS}}=(P_\mu)^2+(M_{\mu\nu})^2$ and
$c_{[2],{\rm spp}}=(P_\mu)^2+({P^*}_i)^2$ 
where $P,M,P^*$ are translation, Lorentz rotation,
boost parts of the Lorentz rotation generators respectively.
In cases of AdS$_2\times$S$^2$ and four-dimensional pp-wave space, 
 $c_{[2],{\rm sAdS}}=c_{[2],{\rm spp}}$.
Analogously to \bref{QQDD} 
the BPS projection operator ${\cal P}_+$ is obtained from 
the right hand side of the anticommutator of supercharges
\bref{SUSYads} in the Majorana supercharge basis \cite{Zhou}
\bea
\left\{ Q_\alpha, Q_\beta \right\}
&\equiv&-2\sqrt{c_{[2]}}{\cal P}_{+,\alpha\beta}~~~\label{BPS},
\eea
and the quadratic Casimir is bounded below
\bea
c_{[2]}\geq K^2+\bar{K}^2=T^2~~~.\label{BPSmass}
\eea
If a D0-brane in a ground state 
has only non-zero value in  $P_0$ component
and is located at the origin, 
the BPS projection operator is given as
\bea
{\cal P}_{+,\alpha\beta}={\bf 1}-\epsilon_{IJ}\left(
\displaystyle\frac{K}{\sqrt{K^2+\bar{K}^2}}
\gamma_0\otimes {\bf 1}
+\displaystyle\frac{\bar{K}}{\sqrt{K^2+\bar{K}^2}}
\gamma_0\gamma\otimes \bar{\gamma}\right)
\eea
where $I=1,2$ for $N=2$ $\gamma_\mu$ and $\bar{\gamma}_{\bar{\mu}}$
are gamma-matrices for AdS$_2$ and S$^2$ spaces and $\gamma=\gamma_0\gamma_1$
and $\bar{\gamma}=\bar{\gamma}_2\bar{\gamma}_3$ \cite{Zhou}.
It can be seen that a D0-brane is a 1/2 supersymmetry preserving BPS state.
For the four dimensional pp-wave background limit
this form stays the same for a D0-brane.
For the flat limit two kappa-coefficients become equal $K=\bar{K}$, 
and reduce into the similar form with \bref{QQDD}.

\vskip 6mm
\section{ Conclusions}\par

We have shown that in curved backgrounds
Noether charges are also modified by the WZ term, 
and anticommutators of fermionic Noether charges  produce a brane charge.
The $p$-form field strength is no longer a trivial element of 
CE cohomology after subtracting the total derivative term from the 
super-invariant WZ term.
The brane charge in the anticommutator of supercharges
is not a center,
but can be absorbed into the bosonic generators
and then the algebra closes.

The commutators of supercharges for a D0-brane in the AdS$_2\times$S$^2$  
and in the pp-wave limit 
are given in \bref{SUSYads} and
\bref{SUSYpp}, and brane charges are given in
\bref{calT} and \bref{calTpp}.
These contributions from the WZ term 
are absorbed  in conjugate momenta,
and the resultant algebras are the same algebra with the original
super-AdS$_2\times$S$^2$/super-pp-wave algebras.
The brane charges have position dependence; $x^\mu~{(\mu=0,1,2,3)}$ dependence in the AdS,
$x^-,x^i~{(i=1,2)}$ dependence in the pp-wave limit and
no position dependence in a flat limit.
In the lightcone gauge, where $P_+\neq 0$ then $Q_+$'s are broken,  
the dynamical supercharges $Q_-$ are preserved 
 only for $x^i=0$ as discussed in \cite{ske}.
In the static  gauge half supersymmetry condition 
is given by the BPS projection ${\cal P}_+$ 
where the square of the quadratic Casimir  is bounded below 
with the value $T=\sqrt{K^2+\bar{K}^2}$. 
It is natural to expect the above properties for general $p$-branes.

It is also important to examine 
general brane configurations. 
This background covariant approach presents 
more general description including a D0-brane,
which is excluded in the lightcone gauge approach
since the lightcone gauge restricts $x^\pm$ to be Neumann boundary condition.
Toward better understanding of D-branes, dualities,
 gauge theory correspondence and quantum theories 
further studies are required.

\indent
\vskip 6mm
{\bf Acknowledgments}\par
We thank Kiyoshi Kamimura, Makoto Sakaguchi 
and Warren Siegel for 
stimulating discussions and 
help of making colour tables.



\vspace{0.5cm}
\begin{thebibliography}{99}
\bibliographystyle{unsrt}
%
\setlength{\itemsep}{0.0in}

\bibitem{Malda} 
L. Maldacena, Adv. Theor. Math. Phys. {\bf 2} (1998) 231, hep-th/9711200;\\
S.S. Gubser, I.R. Klebanov  and A.M. Polyakov, {\it Phys. Lett.} {\bf B248} (1998) 105, hep-th/9802109;
\\
E. Witten, {\it Adv. Theor. Math. Phys.} {\bf 2} (1998) 253, hep-th/9802150.

\bibitem{jose} 
 J. Kowalski-Glikman, {\it Phys. Lett.} {\bf B134} (1984) 194; \\ 
C.M. Hull, {\it Phys. Lett.} {\bf B139} (1984) 39; \\ 
P.T. Chrusciel and  J. Kowalski-Glikman, {\it Phys. Lett.} {\bf B149} (1984) 107;\\  
J. Figueroa-O'farrill and G. Papadopoulos, 
{\it J. High Energy Phys.} {\bf 0108} (2001) 036, hep-th/0105308;\\
M. Blau, J. Figueroa-O'farrill, C. Hull and G. Papadopoulos, 
{\it J. High Energy Phys.} {\bf 0201} (2002) 047,
 hep-th/0110242.

\bibitem{BMN} 
D. Berenstein, J. Maldacena and H. Nastase,
{\it J. High Energy Phys.} {\bf 4}  (2002) 013, hep-th/0202021.

\bibitem{MT} 
R.R. Metsaev and A.A. Tseytlin, {\it Nucl. Phys.} {\bf B533} (1998) 109, hep-th/9805028.

\bibitem{Ram} 
For example,\\
R. Kallosh and J. Rahmfeld, {\it Phys. Lett.} {\bf B443} (1998) 143, hep-th/9808038;\\
J. Rahmfeld and A. Rajaraman, {\it Phys. Rev.} {\bf D60} (1999) 64014,  hep-th/9809164;\\
I. Pesando, {\it J. High Energy Phys.} {\bf 11} (1998) 002, hep-th/9808020; {\it J. High Energy Phys.} {\bf 2} (1999) 007, hep-th/9809145;\\
J. Park and S-J. Rey, {\it J. High Energy Phys.} {\bf 1} (1999) 001, hep-th/9812062.

\bibitem{Zhou}
J-G. Zhou, {\it Nucl. Phys.} {\bf B559} (1999) 92, hep-th/9906013.

\bibitem{Me1}
R.R. Metsaev, {\it Nucl. Phys.} {\bf B625} (2002) 70, hep-th/0112044;\\
 R.R. Metsaev and A.A. Tseytlin,
{\it Phys. Rev.} {\bf D65} (2002) 126004,  hep-th/0202109.

\bibitem{ske}
M. Bill$\acute{\rm o}$ and I. Pesando, {\it Nucl. Phys.} 
{\bf B536} (2002) 121,
hep-th/0203028;\\
A.~Dabholkar and S.~Parvizi,
{\it Nucl. Phys.} {\bf B641} (2002) 223,
hep-th/{0203231}; \\ 
D. Bak,
 hep-th/0204033;  \\
K. Skenderis and M. Taylor,
{\it J. High Energy Phys.} {\bf 6} (2002) 025,
hep-th/0204054; hep-th/0211011; hep-th/0212184;  \\
 P. Lee, J. Park, 
 {\it Phys. Rev.} {\bf D67} (2003) 0206002, hep-th/0203257; \\ 
 P. Bain, P. Messen and M. Zamarkar, 
 ``{\it Supergravity solutions for D-branes in Hpp wave backgrounds}", hep-th/0205106;\\
O. Bergman, M.R. Gaberdiel, M.B. Green,
{\it D-brane interactions in type IIB plane-wave background}, hep-th/
0205183;  \\
Y. Michishita, {\it J. High Energy Phys.} {\bf 10}
(2002) 048,
hep-th/0206131.
 

\bibitem{ppex}
For example;\\
M. Cvetic, H. Lu and C.N. Pope, 
 hep-th/0203082;
{\it Nucl. Phys.} {\bf B644} (2002) 65, hep-th/0203229;  \\
 A. Kumar, R.R. Nayak and Sanjay, {\it Phys. Lett.} {\bf B541} (2002) 183,
   hep-th/0204025; \\ 
H. Singh, {\it M5-branes with 3/8 supersymmetry in pp-wave background},
hep-th/0205020;  \\
M. Alishahiha and A. Kumar, 
{\it Nucl. Phys.} {\bf B542} (2002) 130,
hep-th/0205134;   \\
Y. Hikida and Y. Sugawara, 
{\it J. High Energy Phys.} {\bf 10} (2002) 67,
hep-th/0205200.

 \bibitem{WitOl} 
D. Olive and E. Witten, {\it  Nucl. Phys.} {\bf B78} (1978) 97;\\
E. Witten. {\it Nucl. Phys.} {\bf B443} (1995) 85, hep-th/9503124.

\bibitem{azcTow} J.A. de Azcarraga and P.K. Townsend, Phys. Rev. Lett. {\bf 62} (1989) 2579.

\bibitem{TowMS} P.K. Townsend, {``{\it M theory from its superalgebra}"},
in ``{\it Cargese 1997, Strings, branes and dualities}" 141-177,
hep-th/9507048.

\bibitem{Towdem} P.K. Townsend, 
``{\it P-brane democracy}" in PASCOS/Hopkins 1995:0271-286 (QCD161:J55:1995), 
hep-th/9507048.


\bibitem{azcTow2} J.A. de Azcarraga, J.M. Izquierdo and P.K. Townsend, 
{\it Phys. Lett.} {\bf B267} (1991) 366.

 \bibitem{Berk}
N. Berkovits, M. Bershadsky, T. Hauer, S. Zhukov and B. Zwiebach, 
{\it Nucl. Phys.} {\bf B567} (2000) 61, hep-th/9907200.

\bibitem{RWS} R. Roiban and W. Siegel, {\it J. High Energy Phys.} {\bf 11} (2000) 024, hep-th/0010104.

\bibitem{HKS}  M. Hatsuda, K. Kamimura and M. Sakaguchi, {\it Phys. Rev.} {\bf D62} (2000) 105024, 
hep-th/0007009.

\bibitem{HKAdS} M.Hatsuda and K.Kamimura, {\it Nucl. Phys.} {\bf B611} (2001) 77,
 hep-th/0106202.
 
\bibitem{WZAdS}
M. Hatsuda and M. Sakaguchi, {\it Phys. Rev.} {\bf D 66} (2002) 045020,
 hep-th/0205092.  
 
\bibitem{Sugi}  
K. Sugiyama and K. Yoshida, 
{\it Nucl. Phys.} {\bf B644} (2002) 113, hep-th/0206070;   
{\it Phys. Lett.} {\bf B546} (2002) 143, hep-th/0206132; \\
  S. Hyun, H. Shin, {\it Nucl. Phys.} {\bf B543} (2002) 115, 
hep-th/0206090. 

\bibitem{Hyun2}  
J-H. Park, {\it J. High Energy Phys.} {\bf 10} (2002) 32,
 hep-th/0208161;\\
K.M. Lee, {\it Phys. Lett.} {\bf B549} (2002) 213,
 hep-th/0209009;\\
N. Nakayama, K. Sugiyama and K. Yoshida,
``{\it Ground state of supermembrane on pp wave}", 
 hep-th/0209081;\\
S. Hyun and J-H. Park, {\it J. High Energy Phys.} {\bf 10} (2002) 70,
hep-th/0209219.

\bibitem{KSosp} K. Kamimura and M. Sakaguchi, 
``{\it $osp(1{\mid}32)$ and extensions of super-AdS$_5\times$S$^5$ algebra}",
hep-th/0301083.

\bibitem{HKSPen}  M. Hatsuda, K. Kamimura and M. Sakaguchi, 
{\it Nucl. Phys.} {\bf B632} (2002) 114, hep-th/0202190; 
{\it Nucl. Phys. } {\bf B637} (2002) 168, hep-th/0204002.


\bibitem{Green} M.B. Green, {\it Phys. Lett.} {\bf B223} (1989) 157;\\
W. Siegel, {\it Phys. Rev.} {\bf D50} (1994) 2799, hep-th/9403144.

 \bibitem{HKSpp}  M. Hatsuda, K. Kamimura and M. Sakaguchi,  {\it Nucl. Phys. } {\bf B644} (2002) 40, hep-th/0207157.


\bibitem{D0p} M. Cederwall, A. von Gussich, B.E.W. Nilsson and A. Westerberg,
{\it Nucl. Phys.} {\bf B496} (1997) 163, hep-th/9610148;\\
M. Cederwall, A. von Gussich, B.E.W. Nilsson, P. Sundell  and A. Westerberg,
{\it Nucl. Phys.} {\bf B490} (1997) 179, hep-th/9611159;\\
M. Aganagic, C. Popescu and J.H. Schwarz, 
{\it  Phys. Lett.} {\bf B393} (1997) 311,
 hep-th/9610249;
{\it Nucl. Phys.} {\bf B495} (1997) 99,
 hep-th/9612080; \\
E. Bergshoeff and P.K. Townsend,
{\it Nucl. Phys.} {\bf B490} (1997) 145,
 hep-th/9611173. 


\bibitem{HKD01} M. Hatsuda, K. Kamimura, {\it Nucl. Phys.} {\bf B520} (1998) 493, 
hep-th/9708001. 



\end{thebibliography}
\end{document}